\documentclass[Article]{aastex631}

\usepackage{amsmath,amssymb,amsfonts}
\usepackage{amsthm}
\usepackage{mathrsfs}
\usepackage{graphicx}

\begin{document}

\title{Analysis of the Stellar Occultations During the Unprecedented Long-Duration Flare}

\author[0000-0003-1419-2835]{K. Bicz}
\affiliation{Astronomical Institute, University of Wrocław, Kopernika 11, 51-622 Wrocław, Poland}

\author[0000-0003-1853-2809]{R. Falewicz}
\affiliation{Astronomical Institute, University of Wrocław, Kopernika 11, 51-622 Wrocław, Poland}
\affiliation{University of Wrocław, Centre of Scientific Excellence - Solar and Stellar Activity, Kopernika 11, 51-622 Wrocław, Poland}

\author[0000-0002-5778-2600]{P. Heinzel}
\affiliation{Astronomical Institute, University of Wrocław, Kopernika 11, 51-622 Wrocław, Poland}
\affiliation{Astronomical Institute, Academy of Sciences of the Czech Republic, 25165 Ond\v{r}ejov, Czech Republic}

\author[0000-0002-8581-9386]{M. Pietras}
\affiliation{Astronomical Institute, University of Wrocław, Kopernika 11, 51-622 Wrocław, Poland}

\author[0000-0001-8474-7694]{P. Pre\'s}
\affiliation{Astronomical Institute, University of Wrocław, Kopernika 11, 51-622 Wrocław, Poland}



\begin{abstract}

In strong stellar and solar flares flare loops typically appear during the decay phase,  providing an additional contribution to the flare emission and, possibly, obscuring the flare emission. Super-flares, common in active, cool stars, persist mostly from minutes to several hours and alter the star’s luminosity across the electromagnetic spectrum. Recent observations of a young main-sequence star reveal a distinctive cool loop arcade forming above the flaring region during a 27-hour superflare event, obscuring the region multiple times. Analysis of these occultations enables the estimation of the arcade’s geometry and physical properties. The arcade's size expanded from 0.213 to 0.391 R$_*$ at a speed of approximately 3.5$\,$km/s. The covering structure exhibited a thickness below 12$\,$200$\,$km, with electron densities ranging from 10$^{13}$ to 10$^{14}\,$cm$^{-3}$ and temperatures below 7$\,$600$\,$K, 6$\,$400$\,$K, and 5$\,$077$\,$K for successive occultations. Additionally, the flare's maximum emission temperature has to exceed 12$\,$000$\,$K for the occultations to appear. Comparing these parameters with known values from other stars and the Sun suggests the structure's nature as an arcade of cool flare loops.
For the first time, we present the physical parameters and the reconstructed geometry of the cool flare loops that obscure the flaring region during the gradual phase of a long-duration flare on a star other than the Sun.

\end{abstract}



\section{Introduction}\label{sec:intro}

Across the vast expanse of the Universe, stars display a captivating array of behaviors and phenomena. Star spots, stellar flares, and overall stellar activity offer insights into the dynamic nature of stars \citep{Howard_2019, Roettenbacher_2018}. Stellar activity varies widely, with stars exhibiting diverse levels of magnetic activity. Younger cool stars are typically more active, displaying frequent flares and larger star spots, while older stars generally exhibit less activity \citep{Davenport_2019}.

Similar to sunspots, when visible on the Sun, starspots are dark, cooler regions appearing in the stellar photosphere due to intense magnetic activity. These spots vary in size, shape, and lifetime, indicating strong magnetic fields that impact the stars' luminosity and spectral characteristics \citep{Flores_2022}. The presence of starspots causes periodic modulations in the stellar light curve, which are larger in amplitude if the starspots are larger or darker \citep{Strassmeier_2009}.

Stellar flares represent highly energetic and rapid events occurring during magnetic reconnection in the coronae of stars. Nonpotential magnetic energy is swiftly released or converted into other forms of energy.
The resulting radiation spans the entire electromagnetic spectrum, ranging from gamma rays to radio emissions. In the impulsive phase of flares, beams of nonthermal electrons are accelerated within the solar or stellar corona, streaming along magnetic field lines toward the chromosphere. There, they heat a dense matter near the feet of the magnetic loops. Simultaneously, a substantial amount of energy is emitted throughout the electromagnetic spectrum from the feet of the loops, manifesting itself as a transient variation in the star's emission.

In the previous work (Paper~I - \citet{Bicz_2024}) we presented an analysis of the distribution of starspots on CD-36 3202 and a study of the long-duration event (LDE) that occurred on this star for the time TBJD (TESS Barycentric Julian Data) 1486.93. CD-36 3202 is a magnetically active young star with an age of around 40 Myr \citep{bell_2015}. The star is a K2V-type star with an effective temperature of approximately 4$\,$800 $-$ 5$\,$000~K, radius of $0.80\pm0.05$~$\mathrm{R_\odot}$, and mass of $0.8\pm 0.1$~$\mathrm{M_\odot}$ (MAST catalog\footnote{http://archive.stsci.edu}). On the surface of CD-36 3202 there are present large spots capable of covering a significant portion of its surface (an order of magnitude higher than the sunspots on the Sun) (Paper~I) and superflares that have energies even 800 times higher than the strongest flares on the Sun. LDE analyzed in this paper is so far the longest-lasting flare detected in {\em TESS} (Transiting Exoplanet Survey Satellite) data. We estimated the number of spots on this star and their parameters (temperature, size, longitude, and latitude). For the first time we were able to estimate the flare's location relative to 
the location of the spots. We also noticed periodic dips on the flare light curve which we ascribe to a coronal cloud obscuring the flare area during stellar rotation.
In the current work, we analyze this phenomenon in details. 

In Paper~I, we conducted a preliminary analysis of this new phenomenon, suggesting that it may be attributed to an optically thick structure obscuring the flaring region, like a filament or a cool loop. The loops above the flaring region undergo visible structural changes due to temperature and density evolution \citep{Heinzel_2018, Song_2016, Ashwanden_book}. Flare-loop arcades cover extensive areas, comparable to active regions, as evident in the SDO/AIA Sun images \citep{Liu_2014,Guo_2021}. In stellar flares, these loop arcades can be even more extensive and can significantly affect the stellar continuum flux as suggested by \citet{Heinzel_2018}.



In our current research, we employed an advanced model to analyze occultation by cool flare loops, offering a comprehensive explanation of the observed flare light curve both qualitatively and quantitatively. This modeling breakthrough allowed us, for the first time on another star, to accurately determine the genuine physical and geometric parameters of the loops responsible for obscuring the flaring emission.

 We described in Section \ref{sec:intro} the stellar object on which the analyzed flare occurred. Our flare analysis methods and data availability are described in Section \ref{sec:methods}. In Section \ref{sec:results}, we describe the results. The discussion and conclusions are presented in Section \ref{sec:disc}.
 
\section{Methods and Data Availability}\label{sec:methods}

\subsection{Observational data}
{\em TESS} observed CD-36 3202 (also known as TIC 156758257) with high-resolution photometry at two-minute intervals during its sectors $6–7$ (spanning from December 15, 2018, to February 1, 2019) and sector 61 (covering January 19, 2023, to February 12, 2023). The raw data from {\em TESS} was processed by the Science Processing Operations Center pipeline. The pipeline is a descendant of the {\em Kepler} mission pipeline based at the NASA Ames Research Center \citep{TESScit}. The Fig. \ref{fig:lcall}a shows the light curve of CD-36 3202 corrected for the starspots variability, made in Paper~I, starting from TBJD 1486.886 (equivalent to JD 2458486.886, January 3, 2019, 14:48 UT; Sector 6), capturing a stellar superflare event. 
Before the mentioned event, there was another flare at a TBJD of approximately 1486.85~days. This flare had an energy of $(5.31 \pm 2.03) \times 10^{34}$~erg (see more details in Paper~I). It was demonstrated that this flare was not part of the main event. One possibility is that it was a precursor flare, leading to the destabilization of a significant magnetic structure on the star and triggering massive, long-lasting magnetic reconnection (similar observations were recorded on the Sun, \citet{Mitra2020, Fletcher2011,ZIMOVETS2009680}), which resulted in an approximately 27-hour stellar flare. Alternatively, it could have been a random, unrelated flare with no connection to the mentioned LDE.
The periodic fluctuations in brightness stem from the rotational modulation of the flaring region that was situated at a high latitude of approximately $\sim\!\!69^\circ$ (Paper~I). The energy released during the white-light flare was estimated to be $(3.99 \pm 1.22) \times 10^{35}$~erg (Paper~I). Remarkably, during the gradual phase of the flare, three consecutive distinct occultations occurred, each separated by nearly identical intervals corresponding to the rotational period (Fig. \ref{fig:lcall}a, \ref{fig:lcall}b). These dimming events are on the gradual phase of the longest flare ever observed (so far) lasting about 27h (Paper~I). Other two possible dimmings may be noted at the next two full rotations but not analyzed due to poor signal-to-noise ratio. Notably, the shape of these occultations evolved over time, and they ceased to reappear after the end of the flare. These unprecedented phenomena were never before observed during the gradual phase of flares and underscore the exceptional nature of our dataset, providing unique insights into stellar activity and flare physics.

\pagebreak

\subsection{The cloud model}
In order to model the radiative properties of the white-light loops, we follow here the approach of \citet{Heinzel_2018}. We approximate the loop structure with a cloud model located above the
stellar surface and having the kinetic temperature $T$, electron density $n_e$, and geometrical thickness $D$.
In the wavelength range of {\em TESS}, the continuum emitted by dense (super)-flare loops will be mainly due
to the hydrogen recombination in the Paschen and Brackett continua (with heads at 820.36~nm, and 1458.4~nm,
respectively), while we neglect the hydrogen free-free process
and Thomson scattering on free electrons (see estimates for cool loops by \citet{Jejcic_2018}). 

\subsubsection{Hydrogen recombination continua}

Here we follow the physical description by \citet{Heinzel_2018, Heinzel_2024}.
 The absorption coefficient for bound-free hydrogen transition (i.e. photoionization) from atomic level $i$ is

\begin{equation}\label{eq::1}
\kappa_{\nu}^{\rm bf} = \alpha_{\nu} (n_i - n_i^{*} e^{-h \nu / k T}) \, ,
\end{equation}
where $\alpha_{\nu}$ is the hydrogen photoionisation cross-section

\begin{equation}
\alpha_{\nu} = 2.815 \times 10^{29} g_{\rm bf}(i,\nu) / i^5 / \nu^3
\end{equation} 
with $g_{\rm bf}$ being the Gaunt factor for bound-free opacity. $n_i$ is the non-LTE population of the hydrogen level $i$ from which the
photoionization takes place and $n_i^*$ is its LTE counterpart. The second term in Equation \ref{eq::1} represents the stimulated emission which is normally
treated as a negative absorption in the radiative-transfer equation. 
The spontaneous emission coefficient due to free-bound transition (i.e. recombination) to level $i$ is 

\begin{equation}
\eta_{\nu}^{\rm bf} = \alpha_{\nu} n_i^{*} (1 - e^{-h \nu / k T}) B_{\nu}(T)  \, ,
\end{equation}
where $B_{\nu}(T)$ is the Planck function at kinetic temperature $T$ of the loop. 
The general form of the continuum source function will be

\begin{equation}
S_{bf} = \frac{\eta_{\nu}}{\kappa_{\nu}}= \frac{n_i^{*} (1 - e^{-h \nu / k T})}{n_i - n_i^{*} e^{-h \nu / k T}} B_{\nu}(T)  \, .
\end{equation}
Defining the departure coefficients from LTE as $b_i = n_i/n_i^*$, we can rewrite $S_{bf}$ as 

\begin{equation}
S_{bf} = \frac{1 - e^{-h \nu / k T}}{b_i - e^{-h \nu / k T}}B_{\nu}(T) \, .
\end{equation}
Finally, assuming LTE conditions, i.e. $b_i=1$, we naturally get 
\begin{equation}\label{eq::sf}
S_{bf} = B_{\nu}(T) \, .
\end{equation}
with 
\begin{equation}
n_i^* = n_p n_e \Phi_i (T) \, ,
\end{equation}
where $\Phi_i (T) = 2.0707 \times 10^{-16} 2i^2 e^{h\nu_i/kT}/T^{3/2}$ is the Boltzmann factor
($\nu_i$ is the ionisation frequency pertinent to level $i$).
In what follows we assume a pure hydrogen plasma for which
$n_p n_e=n_e^2$. 
Detailed non-LTE modeling of prominence clouds (e.g. \citet{Gouttebroze_1993}) shows that cool prominences at
low electron densities of the order of 10$^{10}$ cm$^{-3}$ have $b_3$ and $b_4$ factors close to one and, therefore,
we can safely assume the LTE conditions for denser loops.

The total absorption coefficient $\kappa_{\nu}^{tot}$ in the wavelength range of interest (i.e. the {\it TESS} passband) will be
the sum of continuum opacities of the Paschen and Brackett continua and multiplied by $D$ we get the optical
thickness at a given wavelength (frequency)
\begin{equation}
\tau_{\nu} = \kappa_{\nu}^{tot} D \, .
\end{equation}

\begin{figure*}[h!]
\gridline{\fig{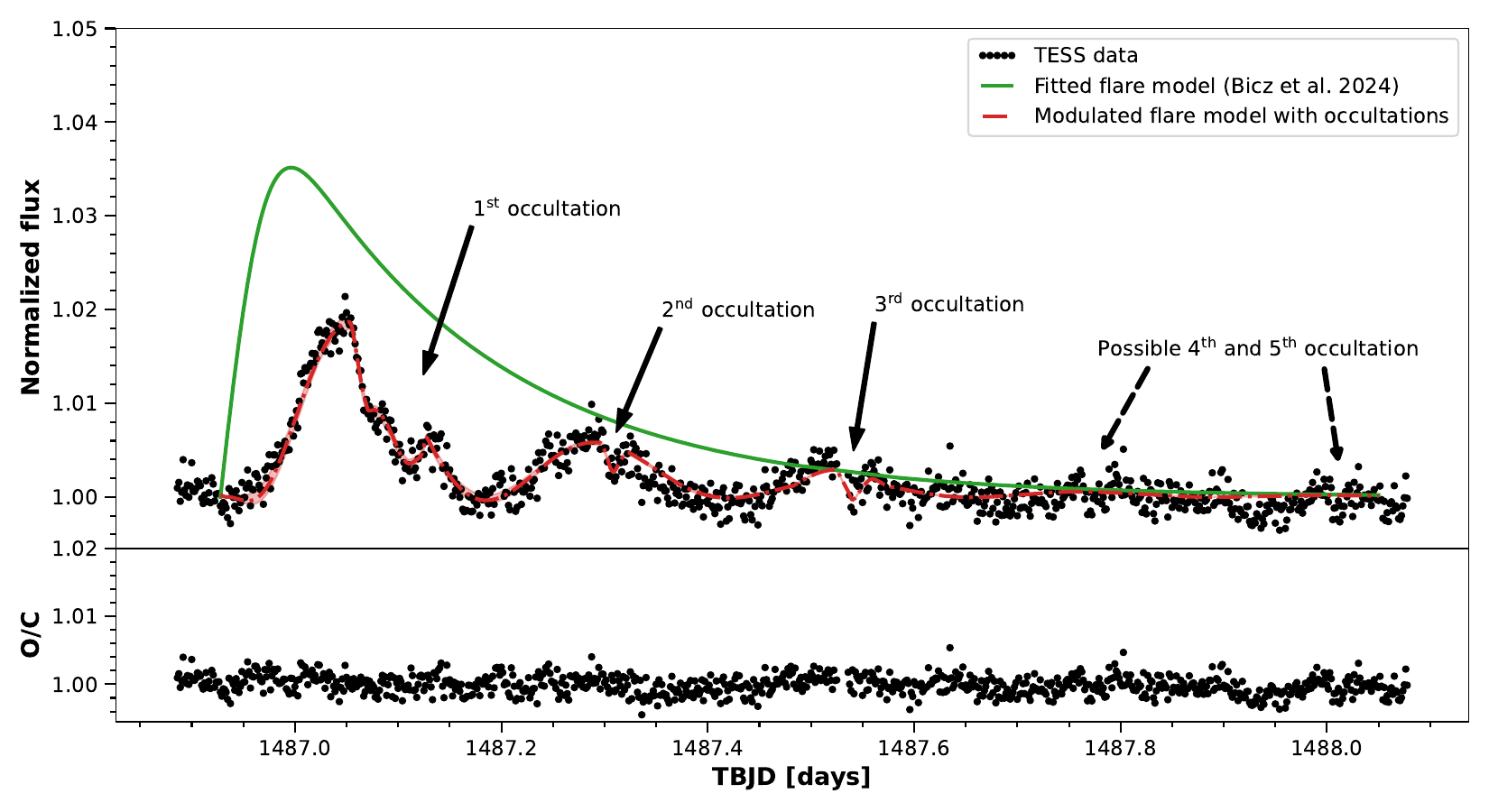}{\textwidth}{(a)}}
\gridline{\fig{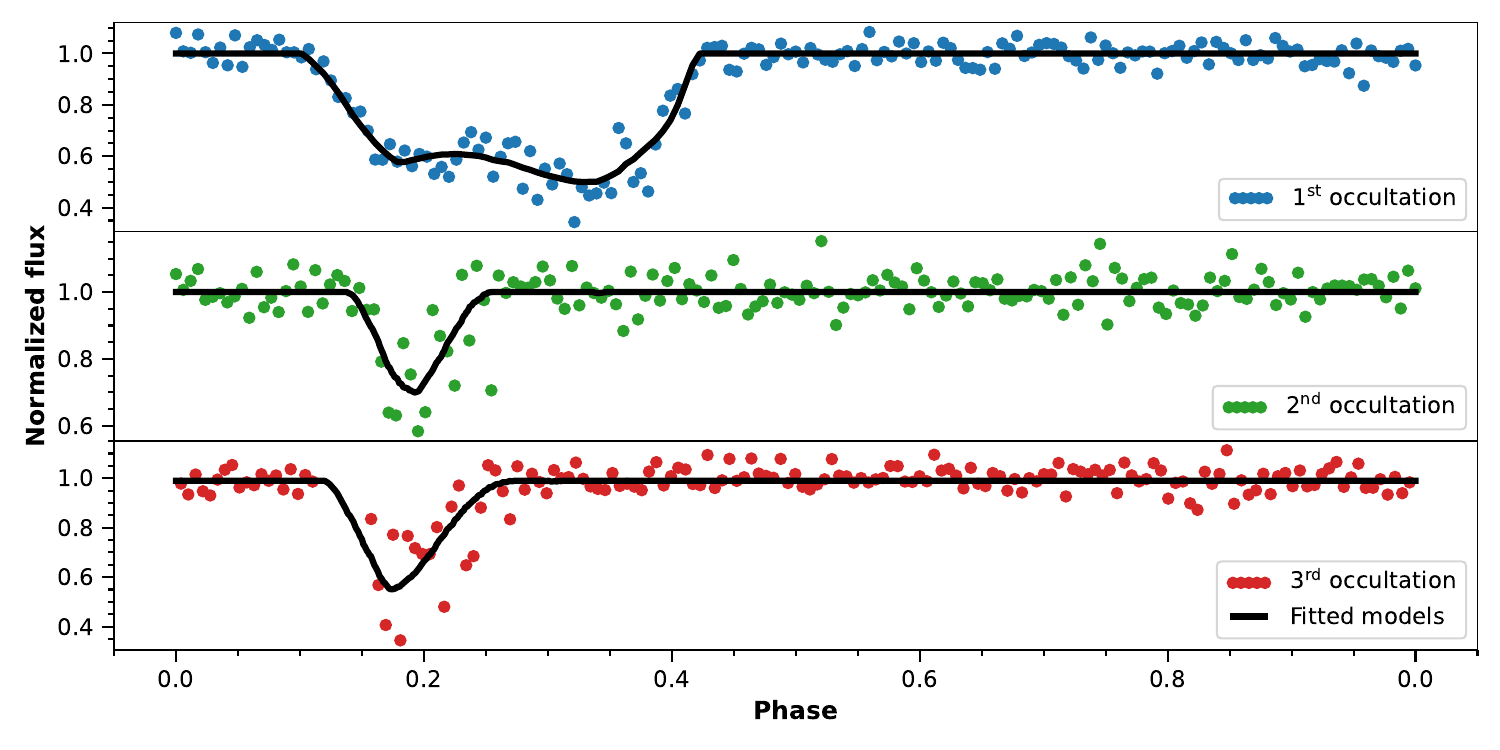}{\textwidth}{(b)}}
\caption{(a) Part of the light curve of CD-36 3202 corrected to the variability caused by the stellar spots. The red curve and red interval mark the modulated flare model with the occultations models and the fit error, respectively. The green curve shows the recreated course of the stellar flare without the effects of foreshortening or stellar rotation from Paper~I. The solid line arrows mark the occultations that were modeled in this paper. The dashed line arrows mark the parts of the light curve that are possible next dimmings. (b) Fragments of the light curve from (a) phased with the rotational period of the star for each of the occultations. First, second, and third occultations are marked with blue, green, and red dots respectively. The black lines present the numerical models for each dimming.}
\label{fig:lcall}
\end{figure*}

\pagebreak
\subsubsection{Radiative transfer model}
In our exploratory model, we assume a cloud with uniform plasma properties specified by the kinetic temperature and electron density, and also
the source function is constant through the slab. The formal solution of the radiative-transfer equation then reads

\begin{equation} \label{eq::rt}
I_{\nu} = I_{\rm bg}  e^{-\tau_{\nu}} + S_{\nu} (1 - e^{-\tau_{\nu}})  \, ,
\end{equation}
where the first term represents the background radiation from the underlying flare ribbons attenuated by the continuum opacity of the loop. The second term is the radiation intensity of the cloud itself.
Note that for moderately thick clouds, the resulting intensity is a mixture of partially penetrating background radiation and the
radiation of the loop itself. We characterize the background ribbon continuum radiation by the Planck function at the temperature
of the chromospheric ribbon $T_f$. Then the free parameters $T$, $T_f$, $n_e$ and $D$ give us the contrast of studied dips.
The contrast $C$ of the dips can be defined as the
\begin{equation} \label{eq::10}
	C = \frac{I - I_{\rm bg}}{I_{\rm bg}}
\end{equation}
Combining the Equations \ref{eq::sf}, \ref{eq::rt}, and the above Equation we get
\begin{equation}
	C = \frac{B_\nu (T_{\rm flare})e^{-\tau} + B_\nu(T_{\rm cloud}) - B_\nu(T_{\rm cloud})e^{-\tau} - B_\nu (T_{\rm flare})}{B_\nu (T_{\rm flare})}
\end{equation}
Which can be rewritten as
\begin{equation}
	C = \left(\frac{B_\nu (T_{\rm cloud})}{B_\nu (T_{\rm flare})} - 1\right) \left(1 - e^{-\tau}\right)
\end{equation}
The above relation for the contrast naturally gives the information about the temperature of the cloud:
\begin{enumerate}
	\item $C = 0 \implies B_\nu (T_{\rm cloud}) = B_\nu (T_{\rm flare}) \iff T_{\rm cloud} = T_{\rm flare}$ - no visible contrast,
	\item $C > 0 \implies B_\nu (T_{\rm cloud}) > B_\nu (T_{\rm flare}) \iff T_{\rm cloud} > T_{\rm flare}$ - cloud in emission,
 	\item $C < 0 \implies B_\nu (T_{\rm cloud}) < B_\nu (T_{\rm flare}) \iff T_{\rm cloud} < T_{\rm flare}$ - cloud in absorption.
\end{enumerate}
If a cloud obscures the flaring ribbons, we expect the cloud's temperature to be lower than that of the flaring ribbon for the occultations to appear.

\subsection{Occultations modelling}

The flare light curve presents three distinct consecutive dimmings separated almost exactly by the rotational period (see Fig. \ref{fig:lcall}b). Another two are noticeable, although their depth is close to the data noise. This behavior may be easily explained by the existence of a co-rotating cloud in the star's atmosphere with measurable optical thickness. The visible change in width occurs between the initial and subsequent dimmings. 
The cloud likely undergoes gradual evolution for approximately 24 hours, after which its impact on the flare's light curve diminishes. Notably, the fourth and fifth occultations remain discernible in the flare's light curve (see Fig. \ref{fig:lcall}a). However, due to the low signal-to-noise ratio during that period, modeling these occultations was not feasible.

To model the obscuring cloud and the light curve we assumed that this structure is part of a loop arcade over the flaring area. We adopted the inclination of the star's rotational axis as $70^\circ$ (Paper~I). The arcade is assumed to consist of $n$ semicircular half-toruses, each representing a loop with a given height $H$ (major-radius of half-torus), thickness $d$ (half-torus cross-section diameter), position angle $\theta$ relative to the local meridian, and the inclination angle $i$ towards the stellar photosphere. We also defined two parameters describing that the arcade may be optically thick only partly, $p$ as the part along the arcade loop, and $l$ describing its center. 
Assuming the loops in the arcade are semicircular half-toruses, and that the overall size of the arcade (measured as $n\,\times\,$the torus cross-section diameter) matches the diameter of the flaring region ($6^\circ 28'$ - Paper I), we can subdivide the arcade into $n$ smaller loops. This is done by dividing the diameter of the flaring region into $n$ equal segments. Instead of having a single semicircular loop with a cross-section diameter equivalent to the flaring region's diameter, we now have $n$ loops, each with a cross-section diameter reduced by a factor of $n$. The thickness of both the loops and the arcade is determined by this cross-section diameter. During the calculations number of loops $n$ varied from one loop up to one hundred loops.

\begin{figure*}[h!]
    \centering 
    \includegraphics[width=\textwidth,height=21.5cm]{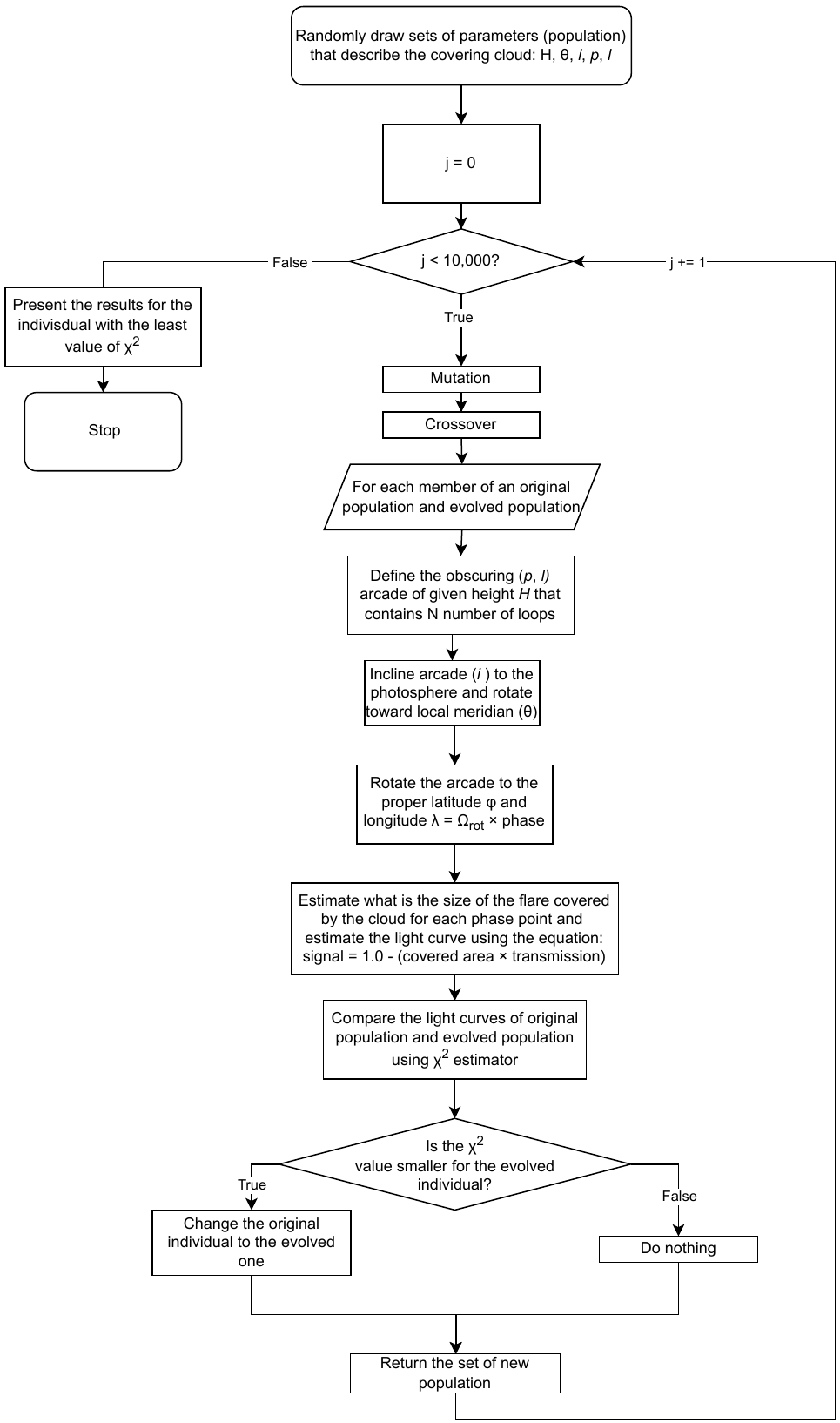}
    \caption{Flow diagram showing the operations made during the light curve modeling.}\label{fig:algorytm}
\end{figure*}
\pagebreak

For each rotational phase, we calculated the position and shape of the flaring area and the obscuring cloud on the plane of sky. 
Then we calculated the size of the obscured part of the flare. For each occultation event, we took that at the light minimum the flare region is totally covered by the cloud. This implies that the effective cloud opacity was determined by the depth of each darkening. The calculated light curves for different sets of parameters were compared with the observations. To find the best fit we used the differential evolution algorithm (\citep{Storn1997, Bicz_2024} and see Fig. \ref{fig:algorytm}). 

\section{Results}\label{sec:results}

In this paper, for the first time, we present the physical parameters and the reconstruction of the geometry of the cool flare loops obscuring the flaring region on a star other than the Sun. The first and second analyzed occultations that occurred in the gradual phase of the LDE started at TBJD 1487.053 and 1487.296, respectively, with depths amounting 50\% and 30\% of the flaring region's signal. A third occultation commenced between TBJD 1487.525 and 1487.536, displaying a depth of 45\% of the flaring region's signal. The exact estimation of the start of the third dimming from the data is difficult due to the short gap in measurements (see Fig. \ref{fig:lcall}b). However, the model of the light curve allowed us to estimate the beginning of the occultation to TBJD 1487.528. In Fig. \ref{fig:lcall}a the dashed arrows mark possible fourth and fifth occultation. Unfortunately, the depth of each of them has a value similar to the noise in the data. This made the modeling of the geometry of the whole structure in latter phases untrustworthy.
Thanks to the geometry modeling, we successfully adjusted the duration times of the dimmings in comparison to those outlined in Paper~I. The first occultation lasted 108 min, the second one 40 min, and the third one 49 min. The best-fit result parameters are presented in Table \ref{tab:lcpar} and illustrated in Fig. \ref{fig:lcall}a and Fig. \ref{fig:lcall}b.

\begin{table*}[h]
\caption{Geometrical parameters of the occulting arcade obtained from the light curve modelling.}\label{tab:lcpar}%
    \begin{tabular}{@{}cccccc@{}}
        \hline
        Occultation \# & Arcade height & Occulting part & Center of the & Inclination & Position angle \\
        & [R$_*$] &  of the arcade & occulting part & [deg] &  [deg] \\
        \hline
        1 & $0.213\pm0.014$ & $0.40\pm0.02$ & $0.22\pm0.02$ & $50\pm1$ & $45\pm2$ \\
        2 & $0.341\pm0.062$ & $0.38\pm0.03$ & $0.29\pm0.05$ & $53\pm2$ & $67\pm3$ \\
        3 & $0.391\pm0.071$ & $0.35\pm0.05$ & $0.32\pm0.06$ & $52\pm2$ & $61\pm5$ \\ 
        \hline
    \end{tabular}
\end{table*}

\begin{figure*}[h]
    \centering
    \includegraphics[width=\textwidth]{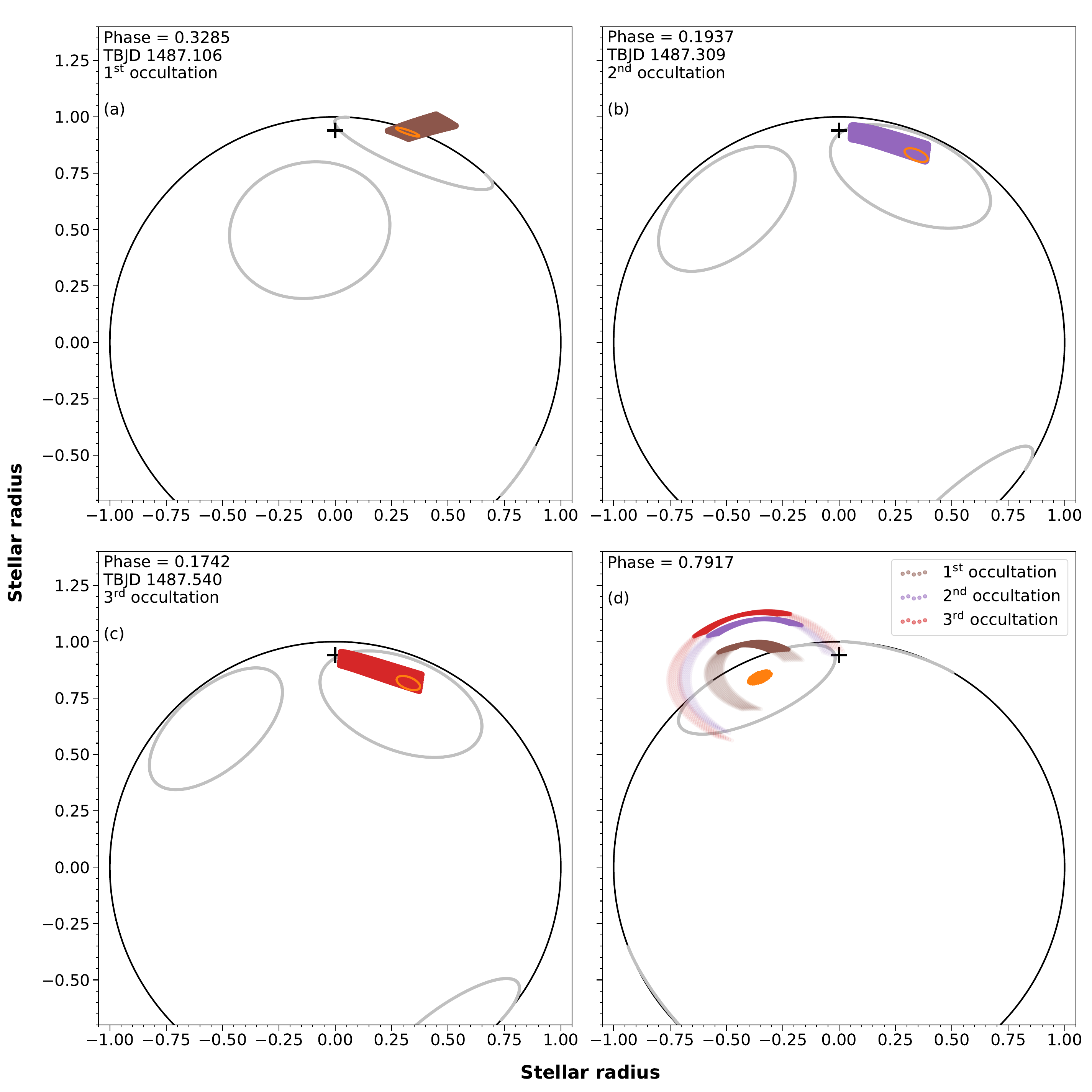}
    \caption{Panels (a), (b), and (c) show the eclipsing part of the loop arcade during the phases of maximum occultation, as indicated by the arrows in Fig. \ref{fig:lcall}a. The orange contours indicate the fully covered flaring region. Panel (d) presents the arcades during selected phase when they do not obscure the flaring region, allowing for size comparisons between the arcades. The brown structure corresponds to the first eclipse, the purple to the second, and the red to the third eclipse. The black cross marks the position of the star's pole. The opaque segments of the arcades dentoe the fragments responsible for the eclipses observed in the light curve, while the transparent segments represent the entire possible arcade. Gray contours denote the starspots on the star from Paper~I, the orange region indicates the flaring region, and the black contours outline the star.}\label{fig:loops} 
    
\end{figure*}

The results show, that for each obscuring event, the cloud is limited to the left side of the flaring area. This comes from the fact, that each darkening happens after the moment when the flare crosses the central meridian and no darkening is registered before that crossing. The whole geometry of the resulting structure compared to the flaring region, starspots from Paper~I, and the star itself can be seen in Fig. \ref{fig:loops}. 

The height of the arcade is slowly rising at the speed of just a few km/s. The speed between the first and the second eclipse is approximately $v_{\mathrm{12}} = 3.5 \pm 1.7$~km/s, between the second and the third, is $v_{\mathrm{23}} = 1.5 \pm 2.6$~km/s, and the average speed is $\bar{v} = 2.4\pm 1.0$~km/s. Such rising speeds are consistent with those found in case of solar flare loops (e.g. see \citet{Jejcic_2018}).

In our analysis, when using the cloud model to obtain the physical parameters of the loop arcade, we assume that there is a moment when the occulting structure completely covers the flaring region. For each of the analyzed dimmings, this happened at the moment when the measured brightness of the flaring region reached its minimum. With the calculated shape of the covering structure, we were able to estimate that the thickness of the structure should be between $3\,800-12\,200$~km for the first occultation, $6\,100-10\,200$~km for the second, and $6\,800-12\,200$~km for the third one to recreate the light curve of the event properly. This corresponds to approximately $5-16$ loops above the flaring region for the first occultation, $6-10$ loops for the second occultation, and $5-9$ loops for the third occultation.

\begin{figure*}[h]
    \centering
    \includegraphics[width=0.9\textwidth]{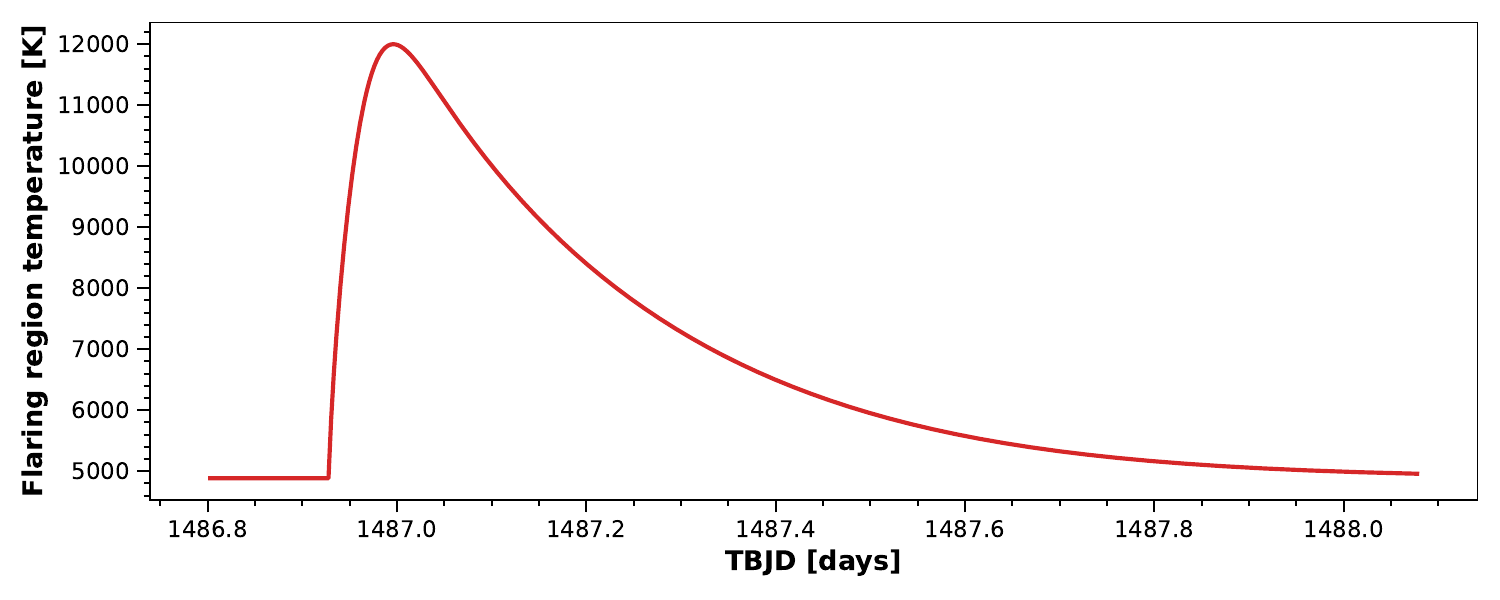}
    \caption{Evolution of the temperature of the flaring region during the whole event.}\label{fig:tempevol}
\end{figure*}

Using the flaring region size from Paper~I, with the assumption that the flaring region size is constant in time, we were able to estimate how the temperature of the flaring region changes using the modified method by us from \citet{Shibayama_2013}. The modification consists in this: instead of using a constant flare temperature and estimating the evolution of the flare area, we assume that the flare size is constant and the temperature changes. Having the size of the flare area from Paper~I, it was possible to estimate how the temperature of the flare area changes (paper presenting this method is in preparation). The evolution of the temperature during the whole event can be seen in Fig. \ref{fig:tempevol}. The temperature of the flaring region for each of the occultations can be seen in Table~\ref{tab:physpar}. Additionally, the cloud model allows us to explain the depth of the eclipses only when the flare’s temperature at the light-curve peak is equal to or higher than 12$\,$000~K. If the temperature was lower than the occultations would not appear.
\pagebreak
\pagebreak

\begin{table*}[h]
\caption{Physical parameters of the occulting arcade obtained from the depth of each occultation.}\label{tab:physpar}%
    \begin{tabular}{@{}cccccc@{}}
        \hline
        Occultation \# & Contrast & Temperature of & Temperature of & Thickness of & $n_e$ \\
        & & the flare [K] & the cloud [K] & the cloud [km] & $[10^{13}\,\mathrm{cm}^{-3}]$ \\
        \hline
        1 & 0.50 & $10\,000$ & $4\,900-7\,600$ & $3\,800-12\,200$ & $1.1 - 28.7$ \\
        2 & 0.30 & $7\,200$  & $4\,900-6\,400$ & $6\,100-10\,200$ & $1.3 - 16.9$ \\
        3 & 0.45 & $6\,000$  & $4\,900-5\,077$ & $6\,800-12\,200$ & $2.6 - 11.8$ \\ 
        \hline
    \end{tabular}
\end{table*}

\begin{figure*}[h]
    \centering
    \includegraphics[width=\textwidth]{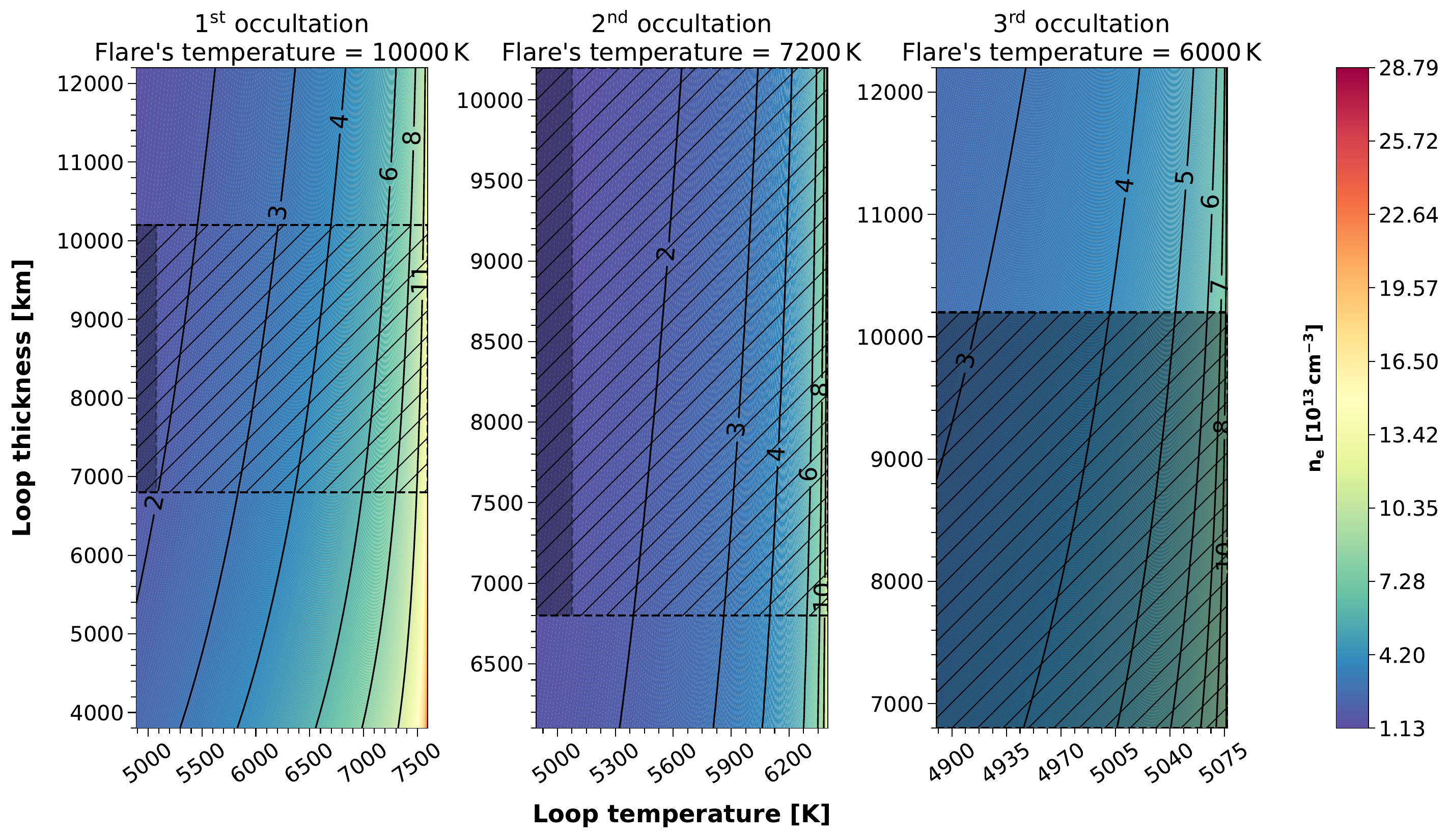}
    \caption{Each of the three panels presents the value of the temperature of the flare region and the relationship between the temperature of the arcade, the thickness of the arcade loops, and the electron density for the first (left panel), second (middle panel), and third eclipse (right panel). The values of electron density are color-coded. Black isolines on each panel represent lines of constant electron density. The surface marked by black oblique lines represents the area where the range of loop thicknesses is the same for each eclipse. The shaded area within these surfaces represents the common range of arcade temperature and thickness.}\label{fig:tt}
\end{figure*}

The temperatures obtained for the flaring region, alongside the estimated thicknesses of the obscuring arcade, allowed us to estimate temperature and density ranges for this structure to reproduce observed occultations. For the initial occultation, temperatures fall within the $4\,900-7\,600$~K range, for the subsequent one between $4\,900-6\,400$~K, and for the third one between $4\,900-5\,077$~K. The estimated temperature ranges of the loop arcade indicate that the arcade must have a significantly lower temperature as compared to the flaring region (by more than 900 K) for the occultations to occur.
The obtained intervals allowed us to estimate the range of electron densities $n_e$ required to achieve the appropriate depth of eclipse (see a second column in Table \ref{tab:physpar}) for each phenomenon.
A summary of the modeling outcomes is provided in Table \ref{tab:physpar} and illustrated in Fig. \ref{fig:tt}.

\section{Discussion}\label{sec:disc}

We analyzed the three occultations on the longest (so far) detected optical stellar flare. It appeared on the star CD-36 3202. The estimated duration times of the eclipses are 108 min, 40 min, and 49 min for the first, second, and third occultation respectively. 
No additional occultations were observed before or after the flare.
We suggest that the covering structure is an arcade of cool, possibly cool flare loops, connected with the active region in which the flare occurred. The areas covered by flare-loop arcades (size of the flaring region is 3000 ppm) are large and comparable to the size of active regions, which is well documented on many images of the Sun from SDO/AIA \citep{Liu_2014,Guo_2021}. In this paper, for the first time, we present the physical parameters and the reconstruction of the geometry of the cool flare loops obscuring the flaring region on a star other than the Sun. This modeling was possible thanks to very good quality photometric data received from the {\em TESS} satellite and the usage of quantitative modeling, which includes all relevant emission and absorption processes. It was possible to reconstruct the geometry of the structure that obscures the flaring region. About 40\% of the arcade was causing occultations. The arcade height increased from $0.213\pm 0.01$~$R_*$ up to $0.391 \pm 0.07$~$R_*$ (which corresponds to the mean expansion speed $\bar{v} = 2.4 \pm 1.0$~km/s). The inclination angle was about 50$^\circ$, and the arcade feet were oriented in range of $45^\circ-67^\circ$ relative to the local meridian.  
The obscuring part of the arcade exhibited a range of temperatures, with minimum values of approximately $4\,900$~K and maximum values reaching up to $7\,600$~K, $6\,400$~K, and $5\,077$~K during the first, second, and third occultations respectively. The thickness of the loops varied, spanning from $3\,800$~km to $12\,200$~km during the first occultation, from $6\,100$~km to $10\,200$~km during the second occultation, and from $6\,800$~km to $12\,200$~km during the third occultation. Required electron densities $n_e$ are then ranging between $1\times10^{13}$ cm$^{-3}$ and $3\times10^{14}$ cm$^{-3}$. There is a specific region of values where the thickness and the temperature of the cloud can be the same for each of the dimmings. This range is $4900 - 5077$~K for the temperature, and $6\,800 - 10\,200$~km for the thickness (black surfaces on Fig. \ref{fig:tt}) which corresponds to the range of $5 - 10$ loops for each occultation.


Our modeling shows that the obscuring structure cannot be a CME as suggested in Paper~I due to the much higher velocities of the CME's \citep{Ravishankar_2020} but can be a prominence or a cool flare loop arcade. The velocity of the whole structure expanding during the flare is approximately $2.4\pm1.0$~km/s. This value perfectly corresponds to the velocity field inside the quiescent prominences that occur on the Sun. 
Similarly, the evolution of the optical cool flare loops above the flaring area has a very similar velocity of a few km/s \citep{Jejcic_2018}.

The arcade that covers the flaring region has a size quite well fitting the size of the starspots on CD-36 3202 (see Fig. \ref{fig:loops}). This is a strong suggestion that the structure itself is connected with the active region and its shape evolution during the whole event is caused by the flare.
During every eclipse only part of the whole arcade obscures the flaring region (see Table \ref{tab:lcpar}). Suppose the higher percentage of the loop was visible above the stellar photosphere then the secondary eclipse would occur during one stellar rotation cycle. This is not visible in the observational light curve (Fig. \ref{fig:lcall}a and Fig. \ref{fig:lcall}b). Partially visible loops are the phenomenon that was also observed on the Sun, where only the fragment of the off-limb loops was visible in the SDO/HMI pseudo-continuum channel during X-class solar flare \citep{Jejcic_2018}. The visible part of the arcade is the one where is the highest electron density compared to other parts of the structure.


The electron density of the covering loops falls within the approximate range of $10^{13} - 10^{14}$~cm$^{-3}$. A relatively high density is needed to absorb enough radiation from the flaring region. A high electron density in the range of $n_e \geq 10^{13}\,$cm$^{- 3}$ is both unquestionably necessary and anticipated, primarily due to the intense evaporative processes occurring during superflares \citep{Heinzel_2018, Kowalski_2015}. Such high coronal densities have indeed been reported for some solar flares \citep{Hiei_1982, Kowalski_2017, Jejcic_2018}. When the loops, due to rotation, appear above the limb with the coronal background, they should be visible in emission, so changing from absorption against the stellar disk into emission against the background corona. Our modeling revealed that the emission from these loops contributes less to the light curve than the observational noise. The contribution from the loops is an order of magnitude lower than the noise in the data. However, it's essential to note that this pertains solely to the phenomenon under our analysis.

The estimated temperatures of the loops fit quite well with the range of plausible temperatures for the cool flare loops \citep{Jejcic_2018}, especially for the first and the second eclipse. The temperatures of the loops for these eclipses reach up to $7\,600$~K and $6\,400$~K for the first and second occultation respectively. For the third eclipse, the maximal temperature of the loops is $5\,077$~K. This is the first indication of such a cool flare loop that ever appeared above the flaring region. The higher values of the arcade temperature would not allow the eclipses to have such depths.
Additionally, the depths of the occultations gave very strict limitations to the temperature of the flare in the moment of maximum emission. It has to be equal to or higher than $12\,000$~K in order to make such an intensity drop possible. For lower temperatures, some of the occultations would not give such a contrast in the light curve. This limitation agrees quite well with the estimations of the flares' peak temperatures presented by \citet{Howard_2020}. For stars of masses higher than 0.42 M$_\odot$ most of the flares had peak temperatures above $12\,000$~K.

Based on the analyzed event, we can confirm that cool plasma arcades can exhibit very solar-like behavior and appear above the starspots or active regions on K-type stars. These arcades were observed for the first time to obscure the flaring region strong enough for the occultations to appear. We suggest that this type of star (partially convective flaring stars) should be the target of further observations (e.g., for the {\em TESS} mission) in order to find more of such events and analyze them in more detail.

\mbox{}\hfill\\
\noindent This work was partially supported by the program "Excellence Initiative - Research University" for years $2020-2026$ for the University of Wrocław, project no. BPIDUB.4610.96.2021.KG. Additional partial support was also provided by the grant of the Czech Funding Agency No.22-30516K and RVO:67985815. The computations were performed using resources provided by Wrocław Networking and Supercomputing Centre (https://wcss.pl), computational grant number 569. The authors appreciate the constructive comments and suggestions from the anonymous referee, which have been very helpful in improving the manuscript. This paper includes data collected by the {\em TESS} mission. NASA's Science Mission Directorate provides funding for the {\em TESS} mission.

\bibliography{bibliography}{}
\bibliographystyle{aasjournal}

\end{document}